\documentclass[lettersize,journal]{IEEEtran}
\usepackage{amsmath,amssymb,amsfonts}
\usepackage{algorithmic}
\usepackage{algorithm}
\usepackage{placeins}
\usepackage{array}
\usepackage[caption=false,font=normalsize,labelfont=sf,textfont=sf]{subfig}
\usepackage{textcomp}
\usepackage{comment}
\newcommand\Tstrut{\rule{0pt}{2.0ex}}         
\newcommand\Bstrut{\rule[-0.9ex]{0pt}{0pt}}   
\usepackage{stfloats}
\usepackage{url}
\usepackage{verbatim}
\usepackage{graphicx}
\usepackage{cite}
\usepackage{color}
\usepackage{multirow}
\usepackage{hyperref}
\usepackage{pifont}
\usepackage{adjustbox}
\usepackage{svg}
\usepackage{xcolor}
\usepackage{tcolorbox}
\usepackage{multirow}
\usepackage{url}
\usepackage{enumerate}
\usepackage{caption}
\usepackage{tcolorbox}
\usepackage{soul}
\usepackage{makecell}
\usepackage[ruled,vlined,algo2e]{algorithm2e}
\begin{document}

\title{Dual-domain Multi-path Self-supervised Diffusion Model for Accelerated MRI Reconstruction}

 \author{Yuxuan Zhang, Jinkui Hao, Bo Zhou
 \thanks{Y. Zhang is with the Department of Radiology, Northwestern University, Chicago, IL, 60611, USA, and the Department of Biomedical Engineering, Huazhong University of Science and Technology, Wuhan, China. J. Hao and B. Zhou are with the Department of Radiology, Northwestern University, Chicago, IL, 60611, USA.
 Corresponding email: bo.zhou@northwestern.edu}
 }

\markboth{Journal of \LaTeX\ Class Files,~Vol.~14, No.~8, August~2021}%
{Shell \MakeLowercase{\textit{et al.}}: A Sample Article Using IEEEtran.cls for IEEE Journals}


\maketitle

\begin{abstract}
Magnetic resonance imaging (MRI) is a vital diagnostic tool, but its inherently long acquisition times reduce clinical efficiency and patient comfort. Recent advancements in deep learning, particularly diffusion models, have improved accelerated MRI reconstruction. However, existing diffusion models' training often relies on fully sampled data, models incur high computational costs, and often lack uncertainty estimation, limiting their clinical applicability. To overcome these challenges, we propose a novel framework, called Dual-domain Multi-path Self-supervised Diffusion Model (DMSM), that integrates a self-supervised dual-domain diffusion model training scheme, a lightweight hybrid attention network for the reconstruction diffusion model, and a multi-path inference strategy, to enhance reconstruction accuracy, efficiency, and explainability. Unlike traditional diffusion-based models, DMSM eliminates the dependency on training from fully sampled data, making it more practical for real-world clinical settings. We evaluated DMSM on two human MRI datasets, demonstrating that it achieves favorable performance over several supervised and self-supervised baselines, particularly in preserving fine anatomical structures and suppressing artifacts under high acceleration factors. Additionally, our model generates uncertainty maps that correlate reasonably well with reconstruction errors, offering valuable clinically interpretable guidance and potentially enhancing diagnostic confidence. The source code and instructions are publicly available at \url{https://github.com/Advanced-AI-in-Medicine-and-Physics-Lab/DMSM}
\end{abstract}

\begin{IEEEkeywords}
Diffusion Model, Accelerated MRI, Self-supervision, Lightweight Model, Uncertainty Estimation
\end{IEEEkeywords}

\section{Introduction}
\IEEEPARstart{M}{agnetic} resonance imaging (MRI) is a widely used imaging technique for disease diagnosis and treatment planning. However, the inherently lengthy acquisition times associated with MRI pose significant challenges in clinical practice. Prolonged scan times can compromise patient comfort, increase motion-related artifacts, and limit patient throughput, hindering the overall efficiency of healthcare systems \cite{vlaardingerbroek2013magnetic}. This is because data samples of an MR image are acquired sequentially in k-space and the speed at which k-space can be traversed is limited by physiological and hardware constraints. To address this, the field has turned to accelerated MRI techniques, which aim to use undersampling of k-space data to reconstruct high-quality MRI.

While undersampling in k-space contravenes the Nyquist-Shannon theorem, leading to aliasing artifacts in image reconstruction with traditional reconstructions, a line of different reconstruction algorithms have been proposed to reconstruct high-quality MRI from accelerated scenarios, including Compress Sensing (CS)-based methods and Deep Learning (DL)-based methods. Compressed sensing (CS)-based reconstruction methods typically employ sparse coefficients in transform domains (e.g., wavelets) alongside application-specific regularizers to iteratively solve ill-posed inverse problems\cite{zhang2015exponential, liang2009accelerating}. However, these iterative sparse optimization approaches tend to produce over-smoothed anatomical structures and may introduce undesirable image artifacts, particularly at high acceleration factors\cite{ravishankar2010mr}. In recent years, DL-based approaches have shown significant improvements in reconstruction accuracy and efficiency over the CS-based approaches, especially under highly accelerated acquisition settings \cite{bernal2019deep}. In early DL-based MRI reconstruction, convolutional neural networks (CNNs) and transformers were widely adopted to mitigate artifacts from undersampled data. For example, Yang et al.~\cite{yang2017admm} proposed ADMM-Net, integrating the ADMM optimization framework with CNNs to bridge model-based and data-driven reconstruction. Schlemper et al.~\cite{schlemper2017deep} introduced a cascaded CNN architecture to progressively refine MRI images, balancing reconstruction speed and quality. In more recent research, Dar et al.\cite{dar2023parallel} propose a novel parallel-stream fusion model (PSFNet) for accelerated MRI reconstruction under limited training data. This research integrates linear scan-specific (SS) priors and nonlinear scan-general (SG) priors via learnable weights to mitigate error propagation and enhance generalization.
Generative Adversarial Networks (GANs) further advanced the field recently by learning implicit image priors~\cite{yang2017dagan,mardani2018deep,quan2018compressed,cole2020unsupervised}. For instance, Yang et al.~\cite{yang2017dagan} developed DAGAN, a conditional GAN incorporating U-Net generators and frequency-domain constraints for real-time reconstruction. Mardani et al.~\cite{quan2018compressed} enhanced structural preservation through cyclic consistency in GAN-based frameworks. However, CNNs and GANs remain limited by local receptive fields and training instability.
To address these issues, transformer-based architectures emerged, leveraging self-attention for global context modeling.    For instance, Huang et al.~\cite{huang2022swin} designed SwinMR with hierarchical Swin transformer blocks to capture long-range dependencies efficiently. Zhou et al.~\cite{zhou2023dsformer} proposed DSFormer, a dual-domain transformer exploiting inter-modality correlations in both image and k-space domains. Despite their advantages, transformers incur high computational costs due to quadratic attention complexity.


Recently, diffusion models have gained significant attention for their ability to produce high-quality, highly detailed images~\cite{yang2023diffusion, ho2020denoising, song2020denoising}. In the accelerated MRI domain, these models have been adapted to address various reconstruction challenges through diverse approaches. 
In methods that leverage diffusion-based priors, Jiang et al.~\cite{gungor2023adaptive} introduce AdaDiff, which employs adaptive diffusion priors through an unconditional generative image prior and a two-phase reconstruction approach, achieving robust and high-quality reconstructions. Similarly, Guan et al.~\cite{guan2024correlated} propose the Correlated and Multi-Frequency Diffusion Model (CM-DM), utilizing multi-frequency priors to constrain noise distributions and improve convergence efficiency, effectively balancing noise suppression and detail preservation.
While these approaches demonstrate the power of incorporating explicit priors, their performance may depend on the quality and generality of the prior assumptions. To address this limitation, recent studies have explored methods that focus instead on optimizing reconstruction directly in the single domain. For example, in the image domain, Ozturkler et al.~\cite{ozturkler2023smrd} develop a regularized 3D diffusion model combined with an optimization method for 3D MRI reconstruction, enhancing image quality and reducing noise without requiring strong prior constraints. Additionally, Geng et al.~\cite{geng2024dp} propose DP-MDM, a detail-preserving framework that employs multiple diffusion models to extract structural and high-frequency features, significantly improving the recovery of fine anatomical details for diagnostic applications. While these methods focus on the image domain reconstruction, the diffusion model also shows promising results in the k-space domain. For example, Ravula et al.~\cite{ravula2023optimizing} explore the theoretical foundations of diffusion models for compressed sensing MRI, proposing a novel optimization strategy that embeds k-space projection operators into the reverse diffusion process. 
Despite the fact that these works have eliminated the requirements of prior assumptions, their methods are generally supervised approaches, meaning that fully-sampled data with long scanning time is necessary for supervised training. Recently, the self-supervised diffusion model has gained increasing attention. Very recently, Mojtaba Safari et al.\cite{safari2025self} proposed a self-supervised adversarial diffusion model, called SSDA-MRI, which integrates an adversarial mapper to capture the conditional distribution in the reverse diffusion process. The method accelerates data
acquisition without requiring fully-sampled datasets. Similarly, Harry et al.\cite{gao2025self} integrated diffusion bridges that learns directly from undersampled measurements by designing a new sub-sampling–based diffusion bridge, which reduces the reliance on fully-sampled k-space data for training. There are also some works that apply the score-based diffusion model, which has recently been applied to MR reconstruction and has shown promising results in reconstruction accuracy and generalization ability\cite{chung2022score}. For instance, Cui et al.\cite{cui2022self} proposed Self-Score, a score-based diffusion model for MRI reconstruction that requires only undersampled data, which first infers the full MR image distribution from undersampled data from Bayesian deep learning, then integrates it with a score-based diffusion model. Similarly, yet more recently, Liu et al.\cite{liu2025score} proposed a self-supervised joint diffusion model (SSJDM) that learns from undersampled k-space measurements using a Bayesian CNN and a score-based diffusion model trained on multi-contrast data, enabling high-quality 3D cardiac MRI reconstruction.

Diffusion models have demonstrated remarkable potential for accelerated MRI reconstruction in generating high-quality, detailed images. However, there are three existing challenges from the prior works in this domain. First, these previous diffusion model methods typically rely on supervised learning, requiring fully-sampled paired datasets that are difficult to obtain in clinical settings~\cite{hyun2018deep,dedmari2018complex,wang2016accelerating,sun2016deep,schlemper2017deep,qin2018convolutional,zhou2020dudornet}. This dependency limits their practical applicability, especially in scenarios where fully-sampled data is scarce or unavailable. Although there are several related self-supervised methods proposed recently~\cite{wang2020neural,safari2025self,gao2025self,cui2022self,liu2025score,yaman2020self,yaman2021zero,zhou2023dsformer,zhou2022dual}, they usually only focus on single-domain reconstruction (e.g., k-space only), therefore resulting in suboptimal reconstructions. In addition, these methods are typically developed for traditional CNN architectures. Second, many previous diffusion models implement transformers and their variants with heavy self-attention computation mechanisms as the backbone of the diffusion model to achieve enhanced reconstruction performance. However, these implementations are typically large and heavy, thus requiring substantial computational and memory costs. Lastly, these methods typically lack mechanisms for uncertainty estimation, which is critical for explainability and clinical decision-making.

\begin{figure*}[htb!]
\centering
\includegraphics[width=0.95\textwidth]{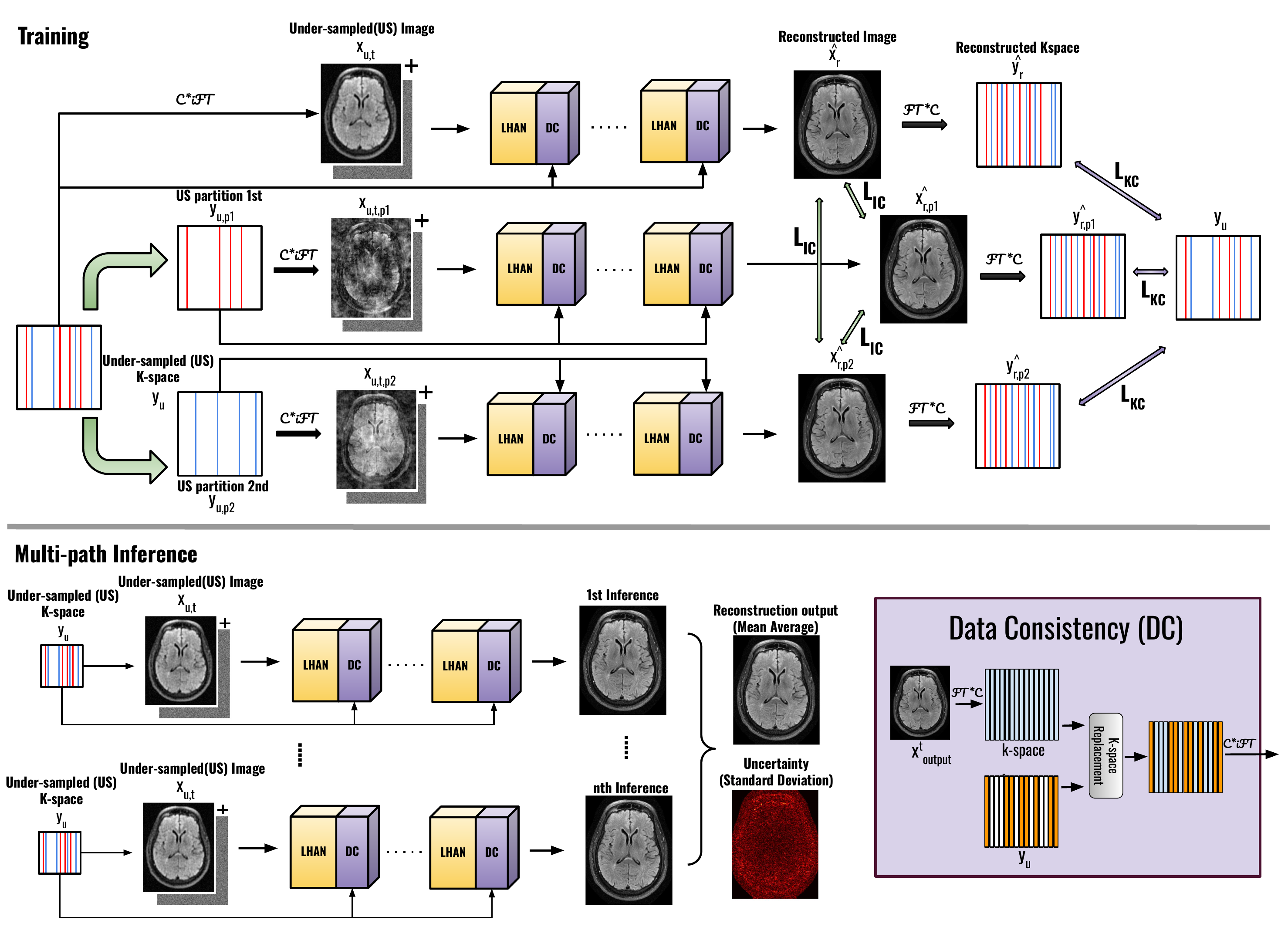}
\caption{Overview of the Dual-Domain Multi-Path Self-Supervised Diffusion Model (DMSM). DMSM applies an unrolled network as a denoiser in the diffusion model, consists of a light-weighted hybrid attention network, followed by a data-consistency layer. During training, the original undersampled k-space $y_u$ is randomly divided into 2 partitions $y_{u,p1}$ and $y_{u,p2}$. The training process takes $y_u,y_{u,p1},y_{u,p2}$ as inputs, and the self-supervised loss is performed on image domain $\mathcal{L}_{IC}$ and k-space domain $\mathcal{L}_{KC}$ for the diffusion model.  Once trained, the diffusion model reconstructs MR images multiple times to get the final output and uncertainty.}
\label{fig:network}
\end{figure*}

To address these limitations, we propose a Dual-domain Multi-path Self-supervised Diffusion Model (DMSM) for accelerated MRI reconstruction with three key innovations. \textbf{Dual-domain Self-supervised Diffusion Model:}  We design a novel dual-domain (k-space and image space) self-supervised training paradigm, addressing the reliance on supervised learning and the suboptimal results of single-domain self-supervised approaches. This enables effective training of the diffusion model using only undersampled k-space measurements, eliminating the need for paired, fully sampled MRI data while leveraging complementary information from both domains. \textbf{Light-weight network for diffusion model:} We develop a lightweight network architecture for the diffusion model to replace computationally heavy Transformer backbones reliant on self-attention. This network incorporates time-step conditioning and a hybrid-attention mechanism designed to efficiently capture essential spatial relationships and features. This innovation achieves high reconstruction quality while significantly reducing computational cost and inference time. \textbf{Multi-path Uncertainty Estimation:} We introduce a novel multi-path sampling strategy within the reverse diffusion process of DMSM to tackle the critical lack of uncertainty quantification in existing methods. This strategy generates multiple plausible reconstructions from the same undersampled data. By analyzing the variance across these paths, we derive pixel-wise uncertainty maps that correlate with reconstruction error, providing crucial explainability for clinical assessment. Our experimental results on two large-scale MRI datasets show that our method can reconstruct high-quality, high-fidelity MRI with training only from under-sampled MRI data and enable uncertainty estimation that correlates
        reasonably well with the true error.

\section{Methods}

The overall architecture of our proposed model is illustrated in Figure \ref{fig:network}. There are three key components in our model: dual-domain self-supervised diffusion model (\ref{method:training}), light-weight hybrid attention network for reconstruction diffusion model (\ref{method:lhan}), and multi-path diffusion model inference strategy (\ref{method:inference}). 

In this section, we also explain the detailed implementation of DMSM (\ref{method:implementation and dataset}), experimental datasets (\ref{method:implementation and dataset}), baseline methods for comparison (\ref{method:metric and baseline}), and evaluation metrics (\ref{method:metric and baseline}).

\subsection{Dual-domain Multi-path Self-supervised Diffusion Model}
We denote $x \in \mathbb{C}^N$ as the MRI image data reconstructed from the acquired k-space data $y \in \mathbb{C}^N$, where x is a vector with the size $N=N_xN_y$, and $N_x$ and $N_y$ are height and width of the image.
In DMSM, we consider the scenario where a conditional diffusion model needs to be trained for accelerated MRI reconstruction. Instead of fully supervised training, DMSM uses a customized dual-domain self-supervised learning strategy only requires under-sampled data. Once trained, the diffusion model can be used for multi-path inference for reconstruction prediction and uncertain estimation. In the following, we discuss the details of our training and inference designs in DMSM. 

\subsubsection{Dual-Domain Self-Supervised Training} \label{method:training}
During training, we employ a partition mask strategy. This approach randomly divides the input under-sampled k-space $y_u$ into two distinct partitions, denoted as $y_{u,{p1}}$ and $y_{u,{p2}}$:
\begin{equation}
    y_{u,{p1}} = M \odot y_u,
\end{equation}
\begin{equation}
    y_{u,{p2}} = (1-M) \odot y_u.
\end{equation}
where $M$ is a binary undersampling mask applied to $y_u$. Then, $y_u$, $y_{u,{p1}}$, and $y_{u,{p2}}$ are inputted into three identical diffusion models where the models' weights are shared during training. 

In each diffusion model, in the forward diffusion process, we gradually add Gaussian noise based on a variance schedule $\beta_t = \beta_1,...,\beta_T$ to $x_0$ (at time step $T=0$) where $x_0$ is an initially reconstructed under-sampled MR image $x_u$ converted from $y_u$:

\begin{equation}
    q(x_t | x_{t-1}) = \mathcal{N} \left( x_t; \sqrt{1 - \beta_t} x_{t-1}, \beta_t I \right)
\end{equation}
Then, our backward diffusion process is applied to gradually remove noise from a pure Gaussian distribution $x_T$ to reconstruct $x_0$ using a denoising reconstruction neural network:
\begin{equation}
        p_\theta (x_{t-1} | x_t) = \mathcal{N} \left( x_{t-1}; \epsilon_\theta (x_t, t), \widetilde{\beta}_t I \right)
\end{equation}
where  $\widetilde{\beta}_t = \frac{1 - \bar{\alpha}_{t-1}}{1 - \bar{\alpha}_t} {\beta}_t$ and $\epsilon_\theta$ represents the denoising neural network parametrized during backward diffusion. And the diffusion model can be trained using the following loss:
\begin{equation}
    \mathcal{L}_{DM} := \mathbb{E}_{t, x_0, \epsilon, y_u} \left[ \left\| \epsilon_t - \epsilon_\theta \left( x_t, t, y_u \right) \right\|^2 \right]
\end{equation}
where $\bar{\alpha}_t = \prod_{m=1}^t \alpha_m$ , $\alpha_t = 1 - \beta_t$ and $\epsilon \sim \mathcal{N}(0, I)$. In this particular task, our reverse diffusion steps are parameterized with an LHAN+DC based backbone network (Figure \ref{fig:network} and to be detailed later). 

Then, our self-supervised losses come from two sources, including image-domain self-supervision and frequency-domain self-supervision for the diffusion model. For the first one, the loss aims to minimize the difference between the diffusion model reconstruction from all three different under-sampled data from the same patient input, thus is defined as: 
\begin{equation}
    \mathcal{L}_{IC}  = \| \hat{x_r} - \hat{x}_{r,p_1} \| + \| \hat{x_r} - \hat{x}_{r,p_2} \| + \| \hat{x}_{r,p_1} - \hat{x}_{r,p_2} \|
\end{equation}
where $\hat{x_r}$, $\hat{x}_{r,p1}$, $\hat{x}_{r,p2}$ are the final outputs of our DMSM reconstruction network $R_{\theta}$:
\begin{equation}
    \hat{x}_{r} = R_\theta(x_{u,t}, y_u, M, C, t)
\end{equation}
\begin{equation}
    \hat{x}_{r,p} = R_\theta(x_{u,t,p}, y_{u,p}, M, C, t)
\end{equation}
where $t$ is the time-index and $C$ is the coil sensitivity map. $x_{u,t}$ is the under-sampled image with scheduled noise added, and $x_{u,t,p}$ is the one from the partitions:
\begin{equation}
    x_{u,t}= \sqrt{\bar{\alpha}_t} x_u + \sqrt{1 - \bar{\alpha}_t} \epsilon
\end{equation}

For the frequency domain self-supervision, the loss is formulated as:
\begin{equation}
    \mathcal{L}_{KC}  = \| \hat{y_r} - {y_{u}} \| + \| \hat{y}_{r,p_1} - y_u \| + \| \hat{y}_{r,p_2} - y_u \|
\end{equation}
where $\hat{y}_{r,p_1}$ and $\hat{y}_{r,p_1}$ are reconstructed k-space from different partitions (i.e. $\hat{x}_{r,p1}$ and $\hat{x}_{r,p2}$). 

Combining all the loss objectives above, our final training loss function can be formulated as:
\begin{equation}
    \mathcal{L}  = \lambda_{IC} * \mathcal{L}_{IC}  + \lambda_{KC} * \mathcal{L}_{KC} + 3*\mathcal{L}_{DM} 
\end{equation}
where $\lambda_{IC}=1$ and $\lambda_{KC}=5$ were set empirically to achieve stable training process.

\begin{algorithm}[ht]
\caption{DMSM Training Scheme}
\KwIn{%
    Batch $\mathcal{B} = \{\mathbf{y}_u, \mathbf{M}, C, t\}$ 
    
    \tcp*[l]{$\mathbf{y_u}$: undersampled k-space, $\mathbf{M}$: partition mask , $C$: coil sensitivity map, $t$: time-index features}
}
\KwOut{Total loss $\mathcal{L}$}

\vspace{0.08cm}
\textbf{Partition k-space:} 

$\mathbf{y}_{u,p1} \leftarrow \mathbf{M} \odot \mathbf{y}_u$, \quad
$\mathbf{y}_{u,p2} \leftarrow (1-\mathbf{M}) \odot \mathbf{y}_u$ 

\vspace{0.08cm}
\textbf{Convert to image domain:}

$\mathbf{x}_u \leftarrow {C}*\mathrm{iFT}(\mathbf{y}_u)$, 

$\mathbf{x}_{u,p1} \leftarrow {C}*\mathrm{iFT}(\mathbf{y}_{u,p1})$, 
$\mathbf{x}_{u,p2} \leftarrow {C}*\mathrm{iFT}(\mathbf{y}_{u,p2})$ 

\vspace{0.08cm}
\textbf{Forward process}

$\alpha_t \leftarrow 1 - \beta_t$,
$\bar{\alpha}_t \leftarrow \prod_{m=1}^{t} \alpha_m$ 

Sample $\epsilon \sim \mathcal{N}(0,I)$ 

$x_t \leftarrow \sqrt{\bar{\alpha}_t} x_{t-1} + \sqrt{1 - \bar{\alpha}_t} \epsilon$ 

$q(x_t|x_{t-1}) = \mathcal{N}\left(x_t; \sqrt{1 - \beta_t} x_{t-1}, \beta_t I\right)$

\vspace{0.08cm}
\textbf{Backward process} 

$\alpha_t \leftarrow 1 - \beta_t$, $\bar{\alpha}_t \leftarrow \prod_{m=1}^{t} \alpha_m$, $\widetilde{\beta}_t \leftarrow \frac{1 - \bar{\alpha}_{t-1}}{1 - \bar{\alpha}_t}\beta_t$ 

$\mu_{t-1} \leftarrow \epsilon_\theta(x_t, t, y_u)$ 

Sample $z \sim \mathcal{N}(0,I)$ 

$x_{t-1} \leftarrow \mu_{t-1} + \sqrt{\widetilde{\beta}_t} \cdot z$ 

$p_\theta(x_{t-1}|x_t) = \mathcal{N}\left(x_{t-1}; \epsilon_\theta(x_t, t), \widetilde{\beta}_t I\right)$ 

\vspace{0.08cm}
\textbf{Reconstruct via shared networks:} 

$\hat{\mathbf{x}}_o \leftarrow \mathrm{LHAN}(\mathbf{x}_{u,t}, t)$,
$\hat{\mathbf{x}}_r \leftarrow \mathrm{DC}(\mathbf{x}_o, y_u)$

\vspace{0.08cm}
\textbf{Compute k-space reconstructions:} 

$\hat{\mathbf{y}}_r \leftarrow \mathrm{FT}*(C\hat{\mathbf{x}}_r)$,
$\hat{\mathbf{y}}_{r,p1} \leftarrow \mathrm{FT}*(C\hat{\mathbf{x}}_{r,p1})$,
$\hat{\mathbf{y}}_{r,p2} \leftarrow \mathrm{FT}*(C\hat{\mathbf{x}}_{r,p2})$ 

\vspace{0.08cm}
\textbf{Calculate losses:} 

$\mathcal{L}_{DM} \leftarrow \mathbb{E}\|\epsilon - \epsilon_\theta(\mathbf{x}_{u,t},t,\mathbf{y}_u)\|^2$ 

$\mathcal{L}_{IC} \leftarrow \|\hat{\mathbf{x}}_r - \hat{\mathbf{x}}_{r,p1}\|_1 + \|\hat{\mathbf{x}}_r - \hat{\mathbf{x}}_{r,p2}\|_1 + \|\hat{\mathbf{x}}_{r,p1} - \hat{\mathbf{x}}_{r,p2}\|_1$ 

$\mathcal{L}_{KC} \leftarrow \|\hat{\mathbf{y}}_r - \mathbf{y}_u\|_1 + \|\hat{\mathbf{y}}_{r,p1} - \mathbf{y}_u\|_1 + \|\hat{\mathbf{y}}_{r,p2} - \mathbf{y}_u\|_1$

\vspace{0.08cm}
\textbf{Aggregate total loss:}

$\mathcal{L} \leftarrow \lambda_{IC}\mathcal{L}_{IC} + \lambda_{KC}\mathcal{L}_{KC} + 3\mathcal{L}_{DM}$ \tcp*[l]{$\lambda_{IC}=1, \lambda_{KC}=5$}

\vspace{0.08cm}
\Return $\mathcal{L}$
\end{algorithm}

\subsubsection{Multi-Path Inference} \label{method:inference}
Once the diffusion model is trained using the above pipeline, we propose a mult-path inference strategy that inference the trained diffusion model with different random noise initiations, generating multiple outputs for a single test data sample. The averaged output is used for enhanced prediction, and the standard deviation is used for uncertainty estimation. The overall inference procedure is illustrated in Figure \ref{fig:network}. Specifically, the reverse diffusion step at time-index $t$ is given by:
\begin{equation}
    x_{t-1} = R_\theta(x_{t}, y_{t}^\epsilon, M, C, t) + \sigma_{t} z
\end{equation}
\begin{equation}
    y_{t}^\epsilon = \mathcal{F}(\sqrt{\bar{\alpha}_{t}} x_u + \sqrt{1 - \bar{\alpha}_{t}} \epsilon_{low})
\end{equation}
where  $\epsilon_{low} \sim \mathcal{N}(0, 0.1I)$  and  $z \sim \mathcal{N}(0, I)$. Let the $N$ independent outputs from the multiple inference paths be denoted as \( \hat{x}_{1}, \hat{x}_{2}, \dots, \hat{x}_{N} \). The average output \( \hat{x}_{\text{avg}} \) and the standard deviation \( \sigma_{\hat{x}} \) across these outputs can be computed as:
\begin{equation}
    \hat{x}_{{avg}} = \frac{1}{N} \sum_{i=1}^{N} \hat{x}_{i}
\end{equation}
\begin{equation}
    \sigma_{\hat{x}} = \sqrt{\frac{1}{N} \sum_{i=1}^{N} (\hat{x}_{i} - \hat{x}_{{avg}})^2}
\end{equation}
where \( \hat{x}_{{avg}} \) represents the averaged reconstructed image and \( \sigma_{\hat{x}} \) gives the uncertainty (standard deviation) of the outputs. 
\begin{figure}[htb!]
\centering
\includegraphics[width=0.45\textwidth]{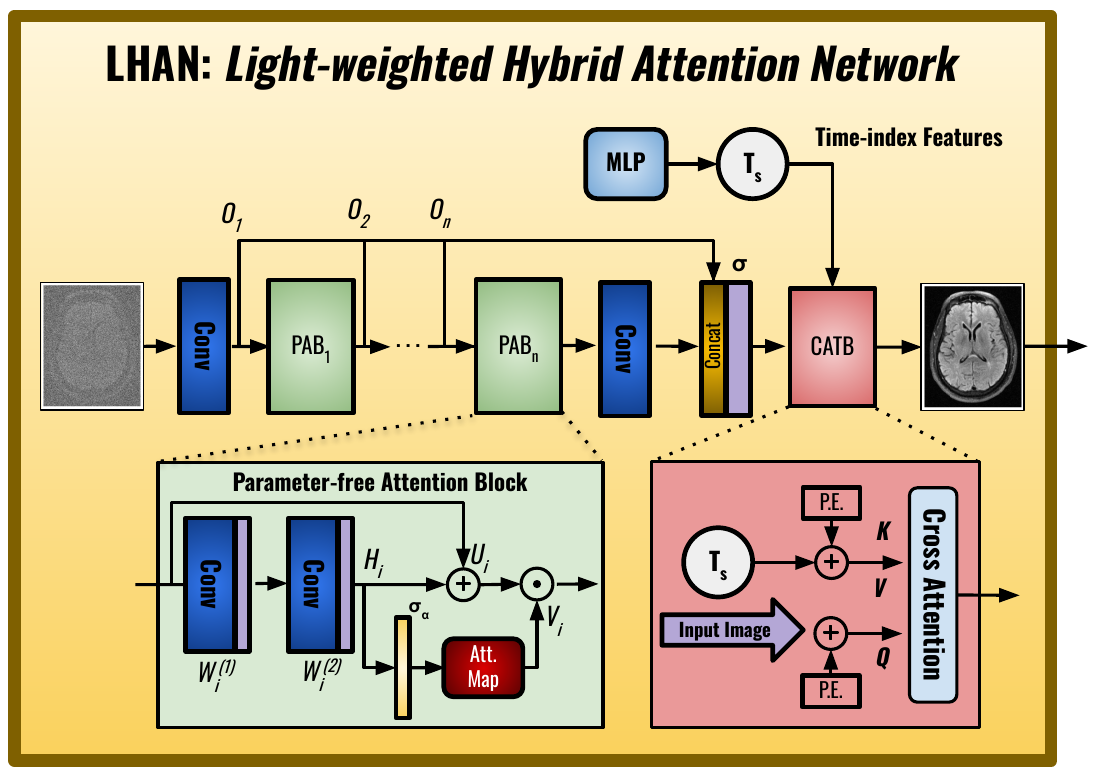}
\caption{Architecture of Light-weighted Hybrid Attention Network (LHAN) in DMSM (Figure \ref{fig:network}). It consists of multiple Parameter-free Attention Blocks (PABs) and a Cross-Attention Transformer Block (CATB). The PABs enable MRI reconstruction feature extraction, while CATB fuses the time index and the extracted feature. The output of LHAN is then input into the DC layer.}
\label{fig:SPCAN}
\end{figure}
\FloatBarrier

\begin{figure*}[htb!]
\centering
\includegraphics[width=0.99\textwidth]{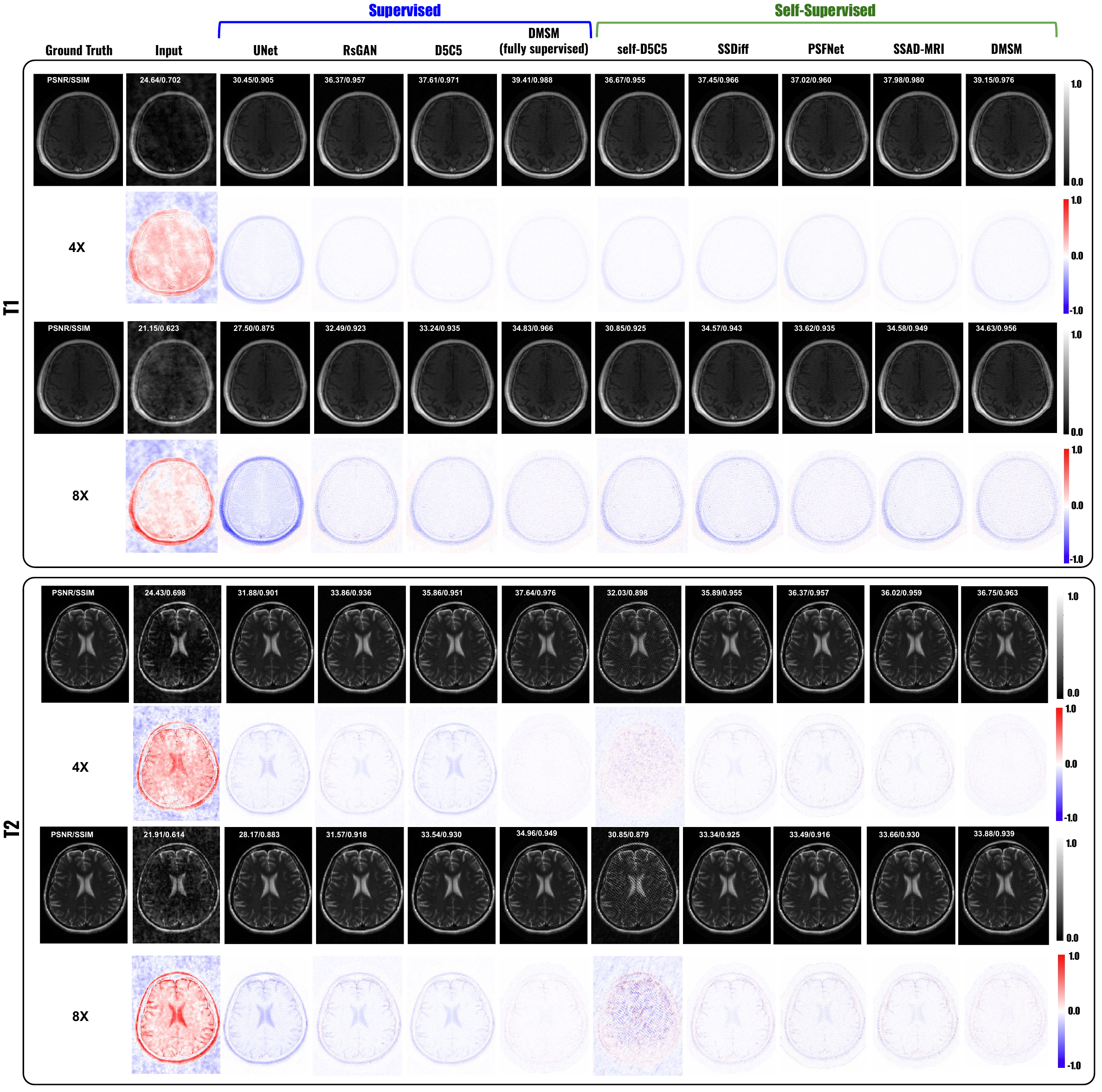}
\caption{{T1 (top) and T2 (bottom) MRI reconstruction results on the fastMRI dataset across all performed baseline methods. Two different acceleration settings (R=4 and R=8) are included. The corresponding error maps are shown right below each reconstruction visualization. Closer to white indicates a better reconstruction compared to the ground truth. PSNR and SSIM values are also reported on the top of the reconstruction results.}}
\label{fig:comp_methods_recon_fastmri}
\end{figure*}

\begin{figure}[htbp]
\centering
\includegraphics[width=0.46\textwidth]{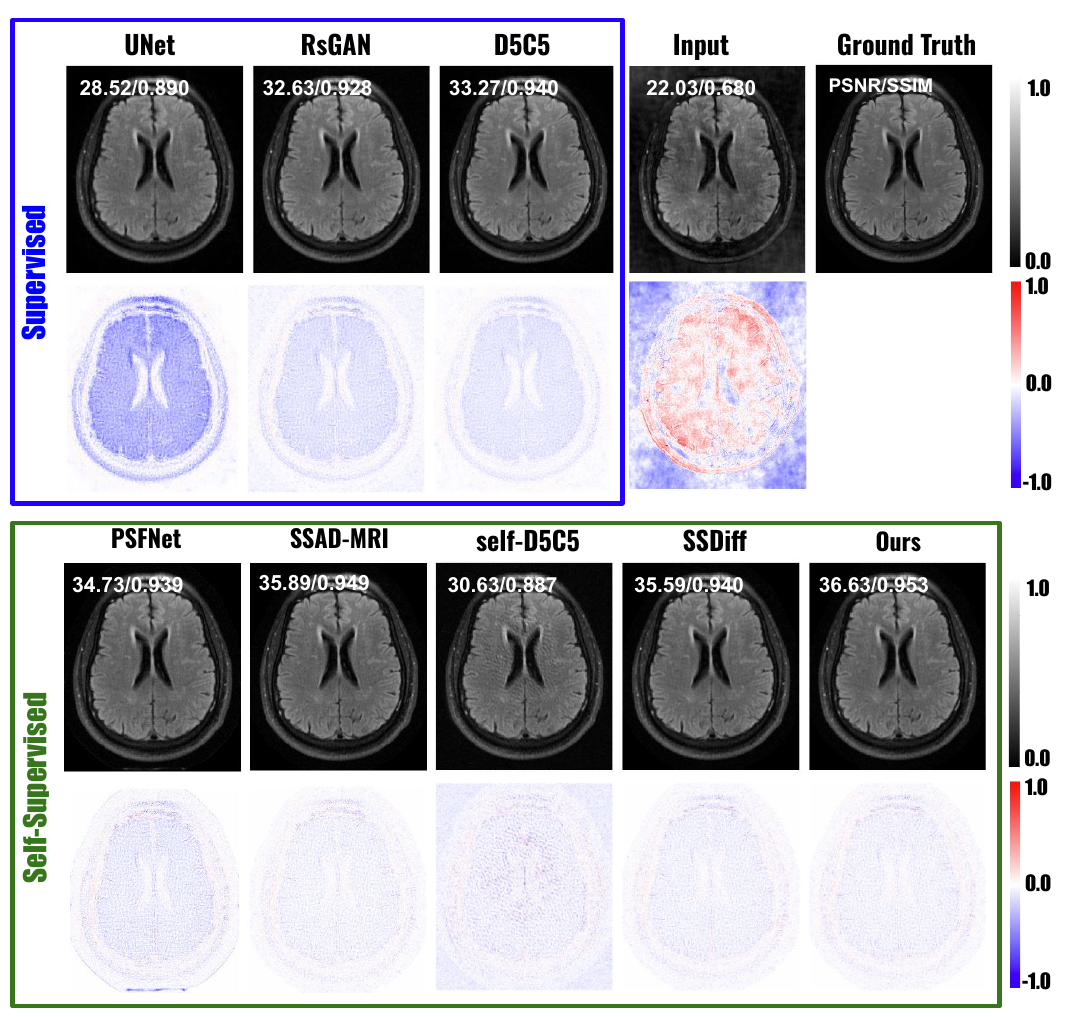}
\caption{{PD MRI reconstruction results on IXI dataset across all performed baseline methods. 4$\times$ acceleration results are presented. The corresponding error maps are shown right below each reconstruction. Closer to white indicates a better reconstruction compared to the ground truth. PSNR and SSIM values are also reported on the top of the reconstruction results.}}
\label{fig:comp_methods_recon_IXI}
\end{figure}

\begin{table*} [!htb]
\centering
\footnotesize
\caption{Quantitative comparisons of reconstruction results on fastMRI dataset. Results on T1, T2, and FLAIR with two highly accelerated rates are reported, including 4$\times$ and 8$\times$. The best results are marked in \textbf{bold}. The fully supervised approach means DMSM is trained in a fully supervised manner with access to paired ground truth. "†" indicates that the difference between DMSM and all supervised and self-supervised baseline methods is significant at $p< 0.00001$ (Bonferroni multiple comparisons-adjusted alpha level) based on the non-parametric Wilcoxon signed rank test.}
\label{tab:compare_method}

\resizebox{\textwidth}{!}{ 
\begin{tabular}{l|c|c|c||c|c|c|c}
\hline
\textbf{T1}     & \multicolumn{3}{c||}{\textbf{4x}} & \multicolumn{3}{c|}{\textbf{8x}} & \textbf{\# Parameters} \Tstrut\Bstrut\\ 
\cline{2-8}
\textbf{Evaluation}   & PSNR/dB & SSIM  & MAE/$*10^{-3}$ & PSNR/dB & SSIM  & MAE/$*10^{-3}$ & \Tstrut\Bstrut\\
\hline
\hline
\text{UNet} & $30.45\pm2.74$ & $0.905\pm0.003$ & $1.45\pm0.72$ & $27.50\pm3.36$ & $0.875\pm0.003$ & $3.60\pm1.50$ & 1.34M \Tstrut\Bstrut\\
\hline
\text{D5C5} & $37.61\pm2.55$ & $0.971\pm0.002$ & $0.23\pm0.47$ & $31.24\pm3.20$ & $0.945\pm0.002$ & $1.70\pm0.44$ & 0.3M \Tstrut\Bstrut\\
\hline
\text{RsGAN} & $36.37\pm2.96$ & $0.957\pm0.002$ & $0.55\pm0.45$ & $30.49\pm3.55$ & $0.913\pm0.002$ & $1.63\pm0.43$ & 11.3M \Tstrut\Bstrut\\
\hline
\text{self-D5C5} & $36.67\pm2.67$ & $0.955\pm0.002$ & $0.47\pm0.39$ & $30.85\pm3.13$ & $0.925\pm0.002$ & $2.56\pm0.73$ & 0.3M \Tstrut\Bstrut\\
\hline
\text{PSFNet} & $37.02\pm2.34$ & $0.960\pm0.002$ & $0.32\pm0.15$ & $33.62\pm3.20$ & $0.935\pm0.002$ & $1.03\pm0.25$ & 8.5M \Tstrut\Bstrut\\
\hline
\text{SSDiff} & $37.45\pm2.30$ & $0.966\pm0.002$ & $0.25\pm0.10$ & $34.57\pm3.14$ & $0.943\pm0.002$ & $0.49\pm0.14$ & 3.3M \Tstrut\Bstrut\\
\hline
\text{SSAD-MRI} & $37.98\pm2.31$ & $0.970\pm0.002$ & $0.20\pm0.11$ & $34.58\pm3.10$ & $0.949\pm0.002$ & $0.44\pm0.21$ & 4.5M \Tstrut\Bstrut\\
\hline
\text{Ours (single-path)} & $38.51\pm3.06$ & $0.972\pm0.002$ & $0.18\pm0.10$ & $33.66\pm2.59$ & $0.948\pm0.002$ & $0.50\pm0.18$ & 0.8M \Tstrut\Bstrut\\
\hline
\text{Ours} & $\mathbf{39.15\pm2.35}^\dagger$ & $\mathbf{0.976\pm0.021}^\dagger$ & $\mathbf{0.16\pm0.10}^\dagger$ & $\mathbf{34.63\pm3.16}^\dagger$ & $\mathbf{0.956\pm0.003}^\dagger$ & $\mathbf{0.35\pm0.10}^\dagger$ & 0.8M \Tstrut\Bstrut\\
\hline
\hline
\text{Fully supervised} & $39.41\pm2.35$ & $0.988\pm0.002$ & $0.12\pm0.04$ & $34.83\pm3.13$ & $0.966\pm0.002$ & $0.29\pm0.10$ & 0.8M \Tstrut\Bstrut\\
\hline
\hline
\end{tabular}
} 

\resizebox{\textwidth}{!}{ 
\begin{tabular}{l|c|c|c||c|c|c|c}
\hline
\textbf{T2}     & \multicolumn{3}{c||}{\textbf{4x}} & \multicolumn{3}{c|}{\textbf{8x}} & \textbf{\# Parameters} \Tstrut\Bstrut\\ 
\cline{2-8}
\textbf{Evaluation}   & PSNR/dB & SSIM  & MAE/$*10^{-3}$ & PSNR/dB & SSIM  & MAE/$*10^{-3}$ & \Tstrut\Bstrut\\
\hline
        
\hline
\text{UNet} & $31.88\pm2.34$ & $0.901\pm0.004$ & $1.2\pm0.37$ & $28.17\pm2.82$ & $0.883\pm0.004$ & $1.01\pm0.40$ & 1.34M \Tstrut\Bstrut\\
\hline
\text{D5C5} & $35.86\pm2.54$ & $0.951\pm0.003$ & $0.28\pm0.11$ & $33.54\pm2.46$ & $0.930\pm0.004$ & $0.52\pm0.21$ & 0.3M \Tstrut\Bstrut\\
\hline
\text{RsGAN} & $33.86\pm2.98$ & $0.936\pm0.003$ & $0.48\pm0.10$ & $31.57\pm3.25$ & $0.918\pm0.003$ & $0.71\pm0.32$ & 8.5M \Tstrut\Bstrut\\
\hline
\text{self-D5C5} & $32.03\pm2.34$ & $0.898\pm0.002$ & $0.55\pm0.25$ & $30.85\pm3.13$ & $0.879\pm0.002$ & $1.63\pm0.43$ & 0.3M \Tstrut\Bstrut\\
\hline
\text{PSFNet} & $36.37\pm2.34$ & $0.957\pm0.002$ & $0.35\pm0.21$ & $33.49\pm2.13$ & $0.916\pm0.002$ & $0.63\pm0.22$ & 11.3M \Tstrut\Bstrut\\
\hline
\text{SSDiff} & $35.89\pm2.40$ & $0.955\pm0.011$ & $0.31\pm0.20$ & $33.34\pm2.46$ & $0.925\pm0.003$ & $0.54\pm0.21$ & 3.3M \Tstrut\Bstrut\\
\hline
\text{SSAD-MRI} & $36.02\pm2.22$ & $0.959\pm0.0015$ & $0.29\pm0.11$ & $33.66\pm2.59$ & $0.930\pm0.002$ & $0.52\pm0.20$ & 4.5M \Tstrut\Bstrut\\
\hline
\text{Ours (single-path)} & $36.74\pm1.93$ & $0.962\pm0.0013$ & $0.25\pm0.15$ & $33.74\pm2.50$ & $0.933\pm0.002$ & $0.50\pm0.20$ & 0.8M \Tstrut\Bstrut\\
\hline
\text{Ours} & $\mathbf{36.75\pm1.93}^\dagger$ & $\mathbf{0.963\pm0.010}^\dagger$ & $\mathbf{0.23\pm0.15}^\dagger$ & $\mathbf{33.88\pm2.25}^\dagger$ & $\mathbf{0.939\pm0.003}^\dagger$ & $\mathbf{0.46\pm0.18}^\dagger$ & 0.8M \Tstrut\Bstrut\\
\hline
\hline
\text{Fully supervised} & $37.64\pm2.24$ & $0.976\pm0.003$ & $0.13\pm0.04$ & $34.96\pm1.97$ & $0.949\pm0.003$ & $0.30\pm0.12$ & 0.8M \Tstrut\Bstrut\\
\hline
\hline
\end{tabular}
} 

\resizebox{\textwidth}{!}{ 
\begin{tabular}{l|c|c|c||c|c|c|c}
\hline
\textbf{FLAIR} & \multicolumn{3}{c||}{\textbf{4x}} & \multicolumn{3}{c|}{\textbf{8x}} & \textbf{\# Parameters} \Tstrut\Bstrut\\
\cline{2-8}
\textbf{Evaluation} & PSNR/dB & SSIM  & MAE/$*10^{-3}$ & PSNR/dB & SSIM  & MAE/$*10^{-3}$ & \Tstrut\Bstrut\\
\hline

\hline
\text{UNet} & $28.69\pm2.78$ & $0.839\pm0.004$ & $1.9\pm0.52$ & $25.80\pm3.58$ & $0.797\pm0.007$ & $2.20\pm0.97$ & 1.34M \Tstrut\Bstrut\\
\hline
\text{D5C5} & $32.96\pm2.75$ & $0.925\pm0.004$ & $0.70\pm0.41$ & $30.85\pm3.54$ & $0.894\pm0.006$ & $0.93\pm0.32$ & 0.3M \Tstrut\Bstrut\\
\hline
\text{RsGAN} & $31.94\pm2.99$ & $0.893\pm0.003$ & $0.89\pm0.31$ & $30.10\pm3.32$ & $0.877\pm0.004$ & $1.50\pm0.51$ & 8.5M \Tstrut\Bstrut\\
\hline
\text{self-D5C5} & $30.63\pm2.34$ & $0.887\pm0.002$ & $0.95\pm0.40$ & $28.97\pm3.13$ & $0.823\pm0.002$ & $1.63\pm0.43$ & 0.3M 
\Tstrut\Bstrut\\
\hline
\text{PSFNet} & $32.95\pm2.68$ & $0.920\pm0.002$ & $0.76\pm0.45$ & $30.49\pm3.13$ & $0.906\pm0.002$ & $1.52\pm0.53$ & 11.3M \Tstrut\Bstrut\\
\hline
\text{SSDiff} & $33.64\pm2.53$ & $0.932\pm0.003$ & $0.68\pm0.25$ & $30.90\pm2.98$ & $0.898\pm0.003$ & $0.81\pm0.32$ & 3.3M 
\Tstrut\Bstrut\\
\hline
\text{SSAD-MRI} & $34.84\pm2.31$ & $0.930\pm0.002$ & $0.66\pm0.19$ & $31.00\pm2.90$ & $0.901\pm0.002$ & $0.79\pm0.30$ & 4.5M \Tstrut\Bstrut\\
\hline
\text{Ours (single-path)} & $33.68\pm2.56$ & $0.929\pm0.002$ & $0.69\pm0.20$ & $31.08\pm2.80$ & $0.905\pm0.003$ & $0.75\pm0.30$ & 0.8M \Tstrut\Bstrut\\
\hline
\text{Ours} & $\mathbf{35.19\pm2.44}^\dagger$ & $\mathbf{0.935\pm0.002}^\dagger$ & $\mathbf{0.59\pm0.19}^\dagger$ & $\mathbf{31.44\pm2.54}^\dagger$ & $\mathbf{0.909\pm0.004}^\dagger$ & $\mathbf{0.71\pm0.31}^\dagger$ & 0.8M \Tstrut\Bstrut\\
\hline
\hline
\text{Fully supervised} & $35.39\pm2.94$ & $0.940\pm0.003$ & $0.60\pm0.21$ & $32.54\pm2.53$ & $0.923\pm0.005$ & $0.55\pm0.21$ & 0.8M \Tstrut\Bstrut\\
\hline
\hline
\end{tabular}
} 

\end{table*}

\begin{table} [htb!]
\centering
\footnotesize
\caption{Quantitative comparisons of reconstructions on IXI dataset. Results on T1, T2, and PD with an accelerated rate equal to 4$\times$ is reported. The best results are marked in \textbf{bold}. The fully supervised approach means DMSM is trained in a fully supervised manner with access to paired ground truth. "†" indicates that the difference between DMSM and all supervised and self-supervised baseline methods are significant at $p< 0.00001$.}
\label{tab:compare_method——IXI}
\resizebox{0.48\textwidth}{!}{
    \begin{tabular}{l|c|c|c}
        \hline
             & \multicolumn{3}{c}{\textbf{T1}}                   \Tstrut\Bstrut\\ \cline{2-4}
        \textbf{Evaluation}   & PSNR/dB & SSIM  & MAE/$*10^{-3}$ \Tstrut\Bstrut\\
        \hline
        
        \hline
\text{UNet} & $32.48\pm3.49$ & $0.934\pm0.004$ & $0.89\pm0.32$ \Tstrut\Bstrut\\
\hline
\text{D5C5} & $38.64\pm3.34$ & $0.978\pm0.003$ & $0.37\pm0.19$ \Tstrut\Bstrut\\
\hline
\text{RsGAN} & $37.84\pm3.48$ & $0.969\pm0.003$ & $0.55\pm0.15$ \Tstrut\Bstrut\\
\hline
\text{self-D5C5} & $37.69\pm3.32$ & $0.972\pm0.003$ & $0.40\pm0.19$ \Tstrut\Bstrut\\
\hline
\text{PSFNet} & $38.95\pm3.55$ & $0.973\pm0.003$ & $0.42\pm0.21$ \Tstrut\Bstrut\\
\hline
\text{SSDiff} & $39.28\pm3.20$ & $0.980\pm0.002$ & $0.31\pm0.07$ \Tstrut\Bstrut\\
\hline
\text{SSAD-MRI} & $39.02\pm3.50$ & $0.979\pm0.002$ & $0.30\pm0.11$ \Tstrut\Bstrut\\
\hline
\text{Ours (single-path)} & $39.10\pm3.70$ & $0.980\pm0.002$ & $0.27\pm0.10$\Tstrut\Bstrut\\
\hline
\text{Ours} & $\mathbf{40.04\pm3.72}^\dagger$ & $\mathbf{0.989\pm0.003}^\dagger$ & $\mathbf{0.19\pm0.04}^\dagger$ \Tstrut\Bstrut\\
\hline
\hline
\text{Fully supervised} & $40.15\pm2.85$ & $0.988\pm0.002$ & $0.12\pm0.04$ \Tstrut\Bstrut\\
\hline
\hline
\end{tabular}
}

\resizebox{0.48\textwidth}{!}{
\begin{tabular}{l|c|c|c}
    \hline
         & \multicolumn{3}{c}{\textbf{T2}}                   \Tstrut\Bstrut\\ \cline{2-4}
    \textbf{Evaluation}   & PSNR/dB & SSIM  & MAE/$*10^{-3}$ \Tstrut\Bstrut\\
    \hline
    
    \hline
\text{UNet} & $30.25\pm3.30$ & $0.910\pm0.004$ & $1.20\pm0.40$ 
\Tstrut\Bstrut\\
\hline
\text{D5C5} & $36.15\pm3.15$ & $0.958\pm0.003$ & $0.48\pm0.20$ 
\Tstrut\Bstrut\\
\hline
\text{RsGAN} & $35.20\pm3.20$ & $0.945\pm0.003$ & $0.60\pm0.25$ 
\Tstrut\Bstrut\\
\hline
\text{self-D5C5} & $32.61\pm3.29$ & $0.922\pm0.005$ & $1.12\pm0.29$ 
\Tstrut\Bstrut\\
\hline
\text{PSFNet} & $35.66\pm3.48$ & $0.953\pm0.003$ & $0.58\pm0.34$ \Tstrut\Bstrut\\
\hline
\text{SSDiff} & $36.98\pm2.95$ & $0.960\pm0.002$ & $0.42\pm0.15$ 
\Tstrut\Bstrut\\
\hline
\text{SSAD-MRI} & $37.15\pm2.94$ & $0.963\pm0.002$ & $0.35\pm0.11$ \Tstrut\Bstrut\\
\hline
\text{Ours (single-path)} & $38.67\pm2.95$ & $0.966\pm0.002$ & $0.29\pm0.12$\Tstrut\Bstrut\\
\hline
\text{Ours} & $\mathbf{38.75\pm2.95}^\dagger$ & $\mathbf{0.970\pm0.003}^\dagger$ & $\mathbf{0.30\pm0.15}^\dagger$ 
\Tstrut\Bstrut\\
\hline
\hline 
\text{Fully supervised} & $39.15\pm2.85$ & $0.975\pm0.003$ & $0.20\pm0.05$ \\
\hline 
\end{tabular}
}

\resizebox{0.48\textwidth}{!}{
\begin{tabular}{l|c|c|c}
\hline
\hline
     & \multicolumn{3}{c}{\textbf{PD}}                   \Tstrut\Bstrut\\ \cline{2-4}
\textbf{Evaluation}   & PSNR/dB & SSIM  & MAE/$*10^{-3}$ \Tstrut\Bstrut\\
\hline
\hline
\text{UNet} & $28.52\pm3.20$ & $0.890\pm0.003$ & $1.60\pm0.65$ 
\Tstrut\Bstrut\\
\hline
\text{D5C5} & $33.27\pm3.55$ & $0.940\pm0.002$ & $0.70\pm0.24$ 
\Tstrut\Bstrut\\
\hline
\text{RsGAN} & $32.63\pm3.13$ & $0.928\pm0.002$ & $0.78\pm0.23$ 
\Tstrut\Bstrut\\
\hline
\text{self-D5C5} & $30.63\pm3.34$ & $0.887\pm0.003$ & $1.07\pm0.29$ 
\Tstrut\Bstrut\\
\hline
\text{PSFNet} & $34.73\pm3.50$ & $0.939\pm0.003$ & $0.69\pm0.23$ \Tstrut\Bstrut\\
\hline
\text{SSDiff} & $35.59\pm3.14$ & $0.940\pm0.002$ & $0.59\pm0.19$ 
\Tstrut\Bstrut\\
\hline
\text{SSAD-MRI} & $35.89\pm3.20$ & $0.949\pm0.002$ & $0.55\pm0.18$ 
\Tstrut\Bstrut\\
\hline
\text{Ours(single-path)} & $36.60\pm3.15$ & $0.95\pm0.002$ & $0.54\pm0.13$\Tstrut\Bstrut\\
\hline
\text{Ours} & $\mathbf{36.63\pm3.16}^\dagger$ & $\mathbf{0.953\pm0.003}^\dagger$ & $\mathbf{0.54\pm0.10}^\dagger$ 
\Tstrut\Bstrut\\
\hline
\hline
\text{Fully supervised} & $36.83\pm3.13$ & $0.964\pm0.002$ & $0.39\pm0.10$ \\
\hline
\end{tabular}
}

\end{table}

\subsection{Backbone Networks in DMSM}\label{method:backbone}
For the backbone of our diffusion model, we develop a Light-weighted Hybrid Attention Network (LHAN), to efficiently generate high-quality reversed images. Then, it is followed by a Data Consistency (DC) Layer to ensure that the reconstructed k-space is consistent with the measured data throughout the reverse process.

\subsubsection{Light-weighted Hybrid Attention Network}\label{method:lhan}
The LHAN architecture consists of multiple Parameter-free Attention Blocks (PABs), and followed by a Cross-Attention Transformer Block (CATB). The PABs are responsible for extracting features with attention to recovering the MRI structures, while the CATB aims to integrate time-indexed embeddings with the PAB features. Specifically, each PAB includes 2 convolution layers for feature extraction and a symmetric activation function for parameter-free attention computation. Given a feature map input $O_{i-1}$, the process can be described as follows:
\begin{equation}
    H_i = \sigma(W^{(2)}_i \star \sigma(W^{(1)}_i \star O_{i-1})), 
\end{equation}
where $W_i^{(1)},W_i^{(2)}$ are two independent convolutional layers, $\star$ represents convolution, and $\sigma$ is the sigmoid activation function. Then, the attention weight $V_i$ is calculated by applying $\sigma_a$ to the initial extracted feature $H_i$ by:
\begin{equation}
    V_i = \sigma_a(H_i),
\end{equation}
where $\sigma_a$ is the symmetric activation function of the original one $\sigma$, which is set to be $Sigmoid(x)-0.5$. The symmetry of $\sigma_{\alpha}$ ensures equal emphasis on positive or negative gradients (e.g., edge directions), while its monotonicity amplifies regions with rich textures, thus generating attention without parameters. Then, the attention $V_i$ is applied to $U_i = O_{i-1} \oplus H_i$ by:
\begin{equation}
    O_i = U_i \odot V_i,
\end{equation}
where $\oplus$ denotes element-wise addition. For the overall process, $O_i$ represents the feature map at each PAB's input or output. After 5 PABs, the final feature map $O_n^t$ is obtained by concatenating features from specific PABs, followed by a convolutional layer. $O_n^t$ then performs as input into CATB, which captures latent features associated with the time-index ${w}_l^t$ obtained from our Time-index Network. 

The Time-index Network takes the time index of the diffusion model, denoted as $t$, as input. It is processed by a Multilayer Perceptron (MLP) network consisting of 12 fully-connected layers, with 32 neurons in each layer. The output of the MLP is a latent vector, which is used to compute cross-attention. Let the time index be represented by $t$, and the input image at time step $t$ be denoted as ${x}_t$. The MLP network processes the time index to produce a latent feature ${w}_l^t$:
\begin{equation}
    {w}_l^t = \text{MLP}(t)
\end{equation}
where ${w}_l^t \in \mathbb{R}^{32}$ is the output latent vector with 32 dimensions, and the MLP function applies a series of transformations through 12 fully connected layers.

The CATB takes the reconstructed image features and time-index latent features as inputs and computes their weighted importance through cross-attention transformer-based network. 
\begin{equation}
    att^t = \operatorname{softmax}\left(\frac{Q\left(O_n^t + P.E.\right) K\left(w_l^t + P.E.\right)^T}{\sqrt{n}}\right) V\left(w_l^t\right),
\end{equation}
\begin{equation}
    O_c^t = \alpha(att) \odot \left(\frac{O_n^t - \mu(O_n^t)}{\sigma(O_n^t)}\right).
\end{equation}
This weighted representation is combined with the original features using a residual connection, 
\begin{equation}
    x_{\text{output}}^t = O_c^t \oplus O_n^t
\end{equation}

\subsubsection{DC Layer}

{The DC Layer is a critical module that enforces k-space data consistency during the reverse diffusion process. After each step, it projects the intermediate estimate $x^{t}{output}$ into measurement space and replaces the sampled k-space positions with the acquired data, ensuring alignment between the evolving reconstruction and scanner measurements. Formally:
        \begin{equation}
            x^{t}_{output} = C*\mathcal{F}^{-1} \left\{ \mathcal{F}(Cx^{t}_{output}) \odot (1 - M) + \mathcal{F}(Cx_u) \odot M \right\}
        \end{equation}
    This physics-driven correction anchors the generative refinement to the acquired measurements, ensuring that reconstructions remain physically valid while the diffusion model infers missing information at unsampled k-space locations.}

\subsection{Data Preparation and Implementation Details}\label{method:implementation and dataset}
We collected two public MRI datasets to evaluate the performance of our proposed method. The first dataset is the fastMRI brain dataset\cite{fastmri_dataset}, which comprises data from 180 subjects. It was partitioned into training, validation, and testing sets with 120, 10, and 50 subjects, respectively. During training, slices from all subjects were randomly shuffled and fed into the model as individual 2D samples. This ensures that our method is exposed to diverse anatomical variations across different slices, thereby enhancing its robustness for single-slice reconstruction. For each subject, it encompasses T1-, T2-, and FLAIR-weighted acquisitions. To mitigate computational complexity, we employed the Generalized Coil Compression (GCC) technique to reduce the number of coils from the original dataset to five, consistent with previous works \cite{zhang2013coil}. The image resolution is $512 \times 512$. The second dataset, IXI\cite{ixi_dataset}, is a simulated single-coil brain MRI dataset, which includes 30 subjects for training, 5 for validation, and 15 for testing. Similar here, each subject contains T1-, T2-, and PD-weighted acquisitions. The image resolution is $256 \times 256$. To simulate the clinical scenario of accelerated MRI, we retrospectively under-sampled the acquisitions using variable-density masks. These masks were generated based on a 2D Gaussian distribution, with variance adjusted to achieve acceleration rates of $R = [4, 8]$, reflecting the undersampling factors used in our experiments. 

We implemented our method in Tensorflow and performed experiments using an NVIDIA GeForce RTX 4090 GPU. We train all models with a batch size of 1 for 500k training steps. The Adam solver was used to optimize our models with $lr=1 \times 10^{-5}$, $\beta_1 = 0.9$, and $\beta_2 = 0.999$. 1000 forward steps are used for the diffusion model.

\subsection{Evaluation Strategies and Baselines}\label{method:metric and baseline}
We evaluate the reconstruction performance using Peak-Signal-to-Noise-Ratio (PSNR), Structural-Similarity-Index (SSIM), and Mean-Absolute-Error (MAE) between reconstructions and the ground truth images (i.e., fully-sampled reconstruction). For the uncertainty estimation evaluation, we use Pearson correlation coefficients (PCC) to calculate the similarity between pixel-wise absolute errors (against fully-sampled ground truth) and our uncertainty estimation from the multi-path inference. For comparative evaluation, we compared our results with several baselines, including supervised baselines such as UNet\cite{wang2016accelerating}, D5C5\cite{kwon2017parallel}, RsGAN\cite{dar2020prior}, and self-supervised methods such as self-D5C5, SSDiff\cite{korkmaz2023self}, PSFNet\cite{dar2023parallel}, SSAD-MRI\cite{safari2025self}. Notably, self-D5C5 is a self-supervised model-based reconstruction method that is trained only on under-sampled MRI data based on SSDU\cite{yaman2020self}. To ensure a fair evaluation of our method, we also compare a single-path version of our model with baseline methods that also operate in a single-path inference setting. The hyperparameters and network architecture are the same as in D5C5. We also compared our DMSM with a fully-supervised strategy version, which is trained with ground truth data. 

\section{Results}
\subsection{Experimental Results}

The visual comparison between our DMSM and other previous methods under different acceleration settings is illustrated in Figure \ref{fig:comp_methods_recon_fastmri} and \ref{fig:comp_methods_recon_IXI}.
In T1-weight reconstruction, as shown in the top panel of Figure \ref{fig:comp_methods_recon_fastmri}, the UNet-based reconstruction exhibited substantial discrepancies from the ground truth. While supervised methods such as RsGAN and D5C5 demonstrated notable visual improvements, their performances diverged in specific aspects. The RsGAN-generated images achieved perceptually enhanced contrast but introduced excessive background noise, leading to degraded quantitative metrics. Replacing D5C5 with a self-supervised paradigm resulted in unstable reconstructions plagued by introducing artifacts and spatial distortions. In contrast, the recently proposed SSDiff and SSAD-MRI methods effectively recovered subtle anatomical structures, yet our DMSM framework achieved superior performance, particularly excelling in preserving edge sharpness and intensity homogeneity.
Under the more challenging R=8 acceleration setting, the limitations of the baseline methods became even more apparent. We noticed that the self-D5C5 failed to produce reliable results, with reconstructions dominated by structural distortions. SSDiff and SSAD-MRI both showed reduced performance compared to the R=4 case, with noticeable intensity drifts in fluid-filled regions in the T2 result. However, DMSM consistently delivered high-quality reconstructions. Similar trends were observed on T2-weighted fastMRI and PD IXI datasets. The visual improvements are more clearly illustrated with the aid of the corresponding error maps, together with the case-specific image quality metrics reported at the top of each image.

The quantitative results for fastMRI and IXI are summarized in Table \ref{tab:compare_method} and Table \ref{tab:compare_method——IXI}. Consistent with our observations from the visualization results, our method achieves the best overall performance across all test samples on average. For instance, under the x4 acceleration setting for T1-weighted imaging, our method attains a PSNR of 39.15 dB, outperforming the previous best method, SSDiff, by approximately 1.7 dB. Additionally, the MAE is reduced by 54.5\%. Under the more challenging x8 acceleration setting, similarly, our method surpasses SSDiff in MAE by 28.6\%. A similar trend is observed in the IXI results. Notably, compared to the fully supervised version of our method, which represents the ceiling performance when paired data is available, our self-supervised DMSM achieves the closest performance. We also report the single-path version of our method, which already demonstrates superior performance over competing approaches, while the full multi-path version achieves the best overall results.

The number of parameters for each model is summarized in the last column of Table \ref{tab:compare_method} to compare their computational burden. As shown, our model achieves the best performance with a compact size of only 0.8M parameters. While D5C5 has a smaller model size (0.3M) due to its use of only cascaded convolutional layers, our approach maintains a model size under 1M while nearly halving the MAE. Compared to the previous best model, SSDiff, we reduce the model size by nearly fivefold while simultaneously enhancing performance.

\begin{figure}[h]
  \centering
  \includegraphics[width=0.46\textwidth]{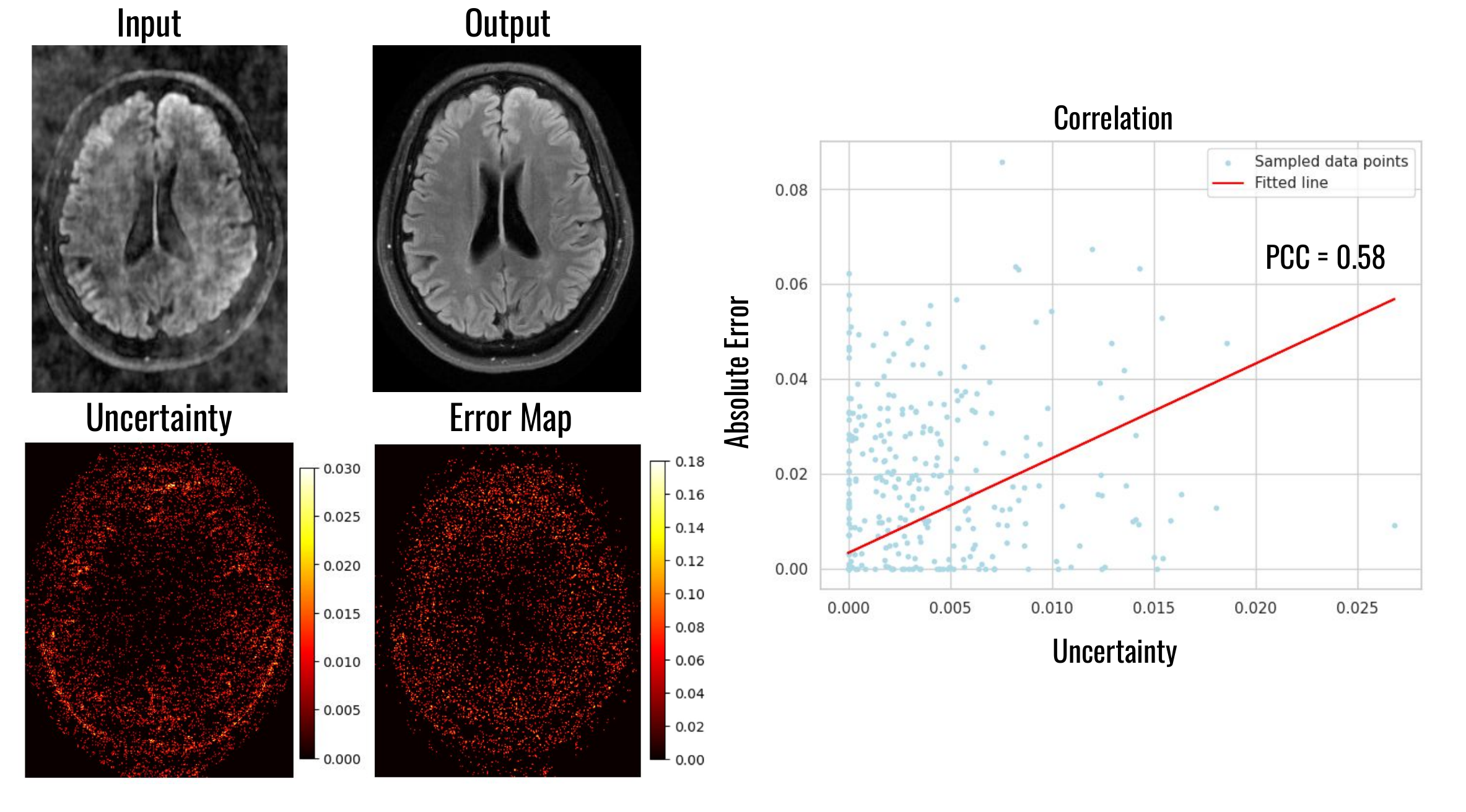}
  \caption{{Uncertainty estimation on the multi-path averaged reconstruction. A result of FLAIR contrast on the fastMRI dataset with 4$\times$ acceleration is shown. Brighter values indicate higher values of variance and bias. The correlation map(right) shows the relationship of absolute error and the uncertainty estimation, with the PCC result attached.}}
  \label{fig:uncertainty}
\end{figure}

Figure \ref{fig:uncertainty} illustrates an example of DMSM's uncertainty estimation and its correlation with reconstruction error (computed using ground truth). As we can see, the input suffers from artifacts due to accelerated acquisition, yet DMSM successfully generates high-quality reconstructions while providing uncertainty estimates that correlate well with the error (PCC = 0.58). The averaged PCC value between the uncertainty estimation and error is $0.44 \pm 0.07$, $0.37 \pm 0.11$, and $0.53 \pm 0.13$ for T1, T2, and FLAIR on fastMRI, respectively. Please note the default number of paths set here is 15. 

\begin{table} [htb!]
\footnotesize
\centering
\caption{\textcolor{black}{Quantitative analysis on uncertainty estimation of DMSM using the fastMRI dataset. Correlation metrics with different numbers of inference paths are reported.}}
\label{tab:uncertainty_quantitative}
    \begin{tabular}{l|c|c|c}
        \hline
        \textbf{Correlation}          & T1            & T2              & FLAIR              \Tstrut\Bstrut\\
        \hline
        \hline
        N = 5          & $0.38\pm0.10$   & $0.35\pm0.11$     & $0.43\pm0.17$     \Tstrut\Bstrut\\
        \hline
        N = 10         & $0.42\pm0.08$   & $0.35\pm0.11$    & $0.54\pm0.13$     \Tstrut\Bstrut\\
        \hline
        N = 15         & $0.44\pm0.07$   & $0.37\pm0.11$     & $0.53\pm0.13$      \Tstrut\Bstrut\\
        \hline
    \end{tabular}
\end{table}

\subsection{Ablation Studies}\label{ablation studies text}

\begin{figure}[htb!]
\centering
\includegraphics[width=0.49\textwidth]{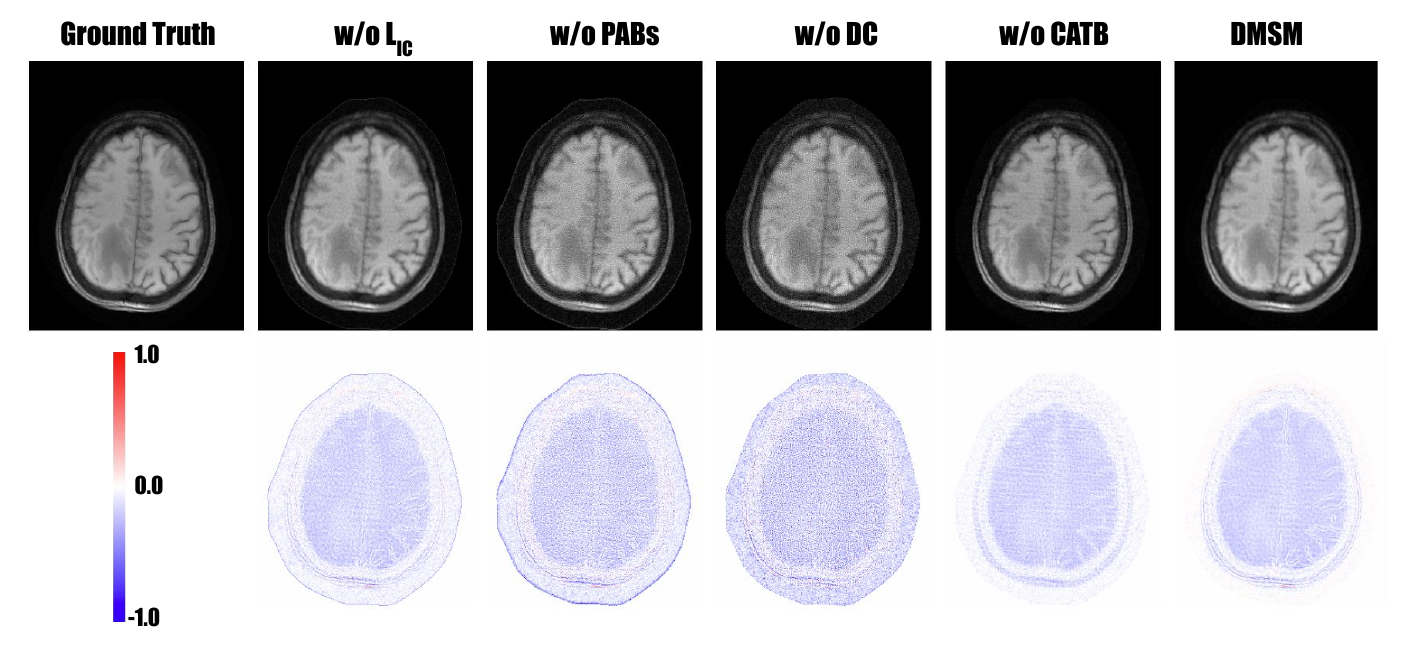}
\caption{{Visualization of ablation studies on DMSM components. The DMSM reconstruction (top row) and the corresponding error map (bottom) row are shown. Results on excluding the image-domain loss (2nd column), LHAN (3rd column), DC layer (4th column) and CATB (5th column) are visualized for a T1 example from fastMRI with 4$\times$ acceleration setting.}}
\label{fig:network_ablation}
\end{figure}

\noindent\textbf{Impact of diffusion self-supervision across different domains:} To assess the effectiveness of our multi-domain self-supervised learning strategy for the diffusion model, we compared DMSM’s performance using only single-domain self-supervision. Figure \ref{fig:network_ablation} provides a visual comparison between models trained with only k-space self-supervision (2nd column) and those incorporating both k-space and image-domain self-supervision (last column). The results demonstrate that multi-domain self-supervision significantly improves reconstruction quality and reduces error. The quantitative comparison in Table \ref{tab:ablation_study} further supports this observation, showing a performance drop from 38.15 dB to 36.39 dB when image-domain self-supervision is omitted. Notably, the model failed to converge when trained with image-domain self-supervision alone, underscoring the necessity of incorporating both domains for effective self-supervision. The results are not reported for image-domain self-supervision for this reason as well.

\begin{table} [htb!]
\footnotesize
\centering
\caption{{Ablation studies of DMSM with or without image domain constraint during training (2nd row), with or without PABs (3rd row), with or without DC layer (4th row), with or without CATB (5th row). T1 on fastMRI dataset with $4\times$ acceleration is used for analysis here.}}
\label{tab:ablation_study}
\resizebox{0.5\textwidth}{!}{%
    \begin{tabular}{l|c|c|c|c}
        \hline
        \textbf{Metrics}          & PSNR/dB            & SSIM              & MAE/$*10^{-3}$       &\# parameters       \Tstrut\Bstrut\\
        \hline
        \hline
        Ours & $39.15\pm2.95$   & $0.976\pm0.021$     & $0.16\pm0.10$   
        & $0.8M$\Tstrut\Bstrut\\
        \hline
        w/o $L_{IC}$         & $36.39\pm3.17$   & $0.958\pm0.037$     & $0.33\pm0.22$   & $0.8M$  \Tstrut\Bstrut\\
        \hline
        w/o  PABs         & $37.65\pm3.10$   & $0.962\pm0.025$     & $0.23\pm0.11$   & $0.6M$   \Tstrut\Bstrut\\
        \hline
        w/o  DC Layer          & $28.82\pm3.94$   & $0.870\pm0.040$     & $1.50\pm0.55$    & $0.8M$  \Tstrut\Bstrut\\
        \hline
        w/o CATB & $38.72\pm2.94$   & $0.966\pm0.020$     & $0.20\pm0.10$
           & $0.4M$   \Tstrut\Bstrut\\
        \hline
    \end{tabular}
    }
\end{table}

\begin{figure}[htb!]
\centering
\includegraphics[width=0.40\textwidth]{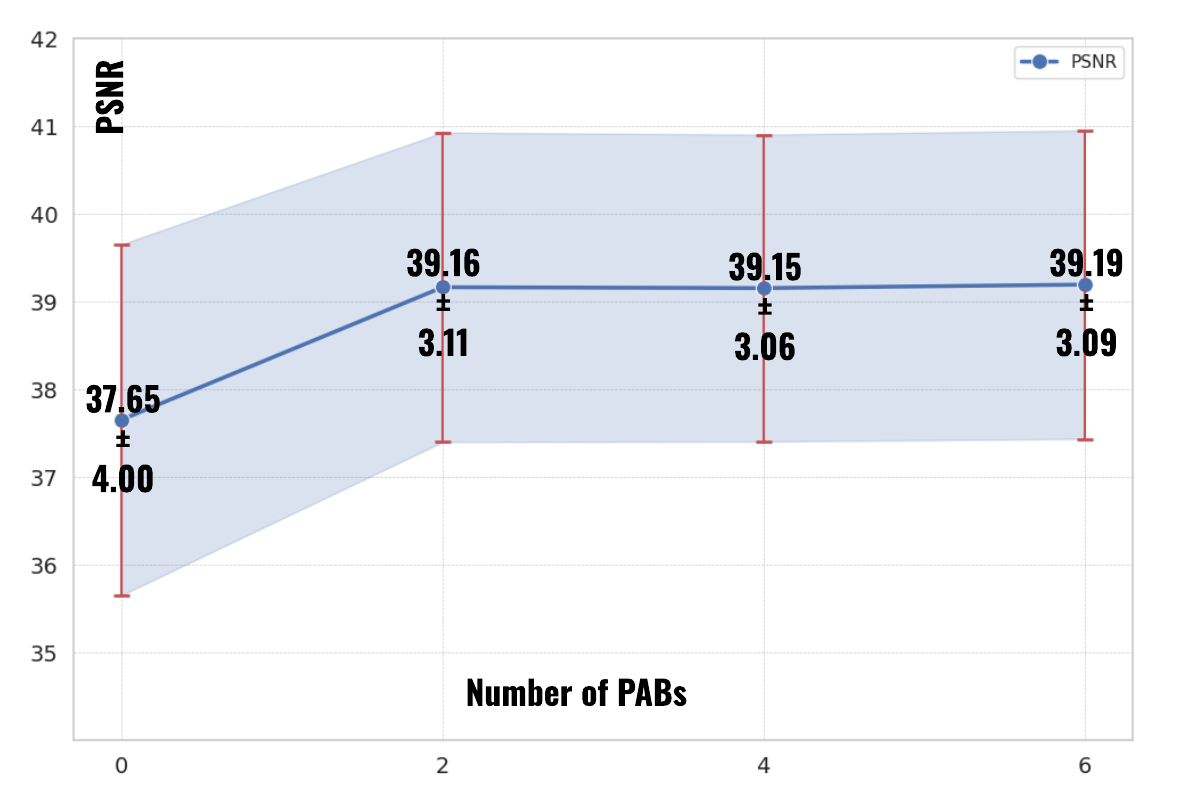}
\caption{{Ablation study on choosing different numbers of PABs inside LHAN. T1 on fastMRI (4$\times$) dataset is used. The PSNR with the corresponding standard deviation is shown. Please note that with the number of PAB equal to 0, only CATB was used. }}
\label{fig:plot_cascade}
\end{figure}

\noindent\textbf{Impact of Sub-Network Structure in DMSM:} The backbone network in DMSM comprises two key modules: LHAN and DC. Here, we extensively evaluate the contribution of each component. Figure \ref{fig:network_ablation} presents a visual comparison between reconstructions without PAB in LHAN (3rd column) and with PAB in LHAN (last column). As shown, PAB effectively extracts features using parameter-free attention, significantly reducing reconstruction error. Table \ref{tab:ablation_study} and Figure \ref{fig:plot_cascade} provide a detailed quantitative analysis of different numbers of PABs in LHAN. Performance is suboptimal without PAB (i.e., number of PAB = 0) but improves and converges when the number of PAB reaches to about 4. Additionally, we examine the impact of DC in the diffusion process. Both the visual examples in Figure \ref{fig:network_ablation} and the quantitative results in Table \ref{tab:ablation_study} show that removing DC severely degrades performance, decreasing from 39.15 dB to 28.82 dB. This highlights the critical role of data consistency in ensuring effective diffusion-based reconstruction.
{We further conducted an ablation study to evaluate the impact of removing the CATB component from our LHAN architecture. Quantitative results indicate a marginal performance decrease, while the parameter count significantly reduces from 0.8M to 0.5M. We further analyze this performance-parameter trade-off in the Discussion section.}

\noindent\textbf{Impact of Multi-Path Inference Strategy:} The multi-path inference strategy simultaneously enhances reconstruction performance and enables uncertainty estimation. Here, we further investigate the effect of varying the number of inference paths in DMSM. Table \ref{tab:multi-number-psnr} summarizes the reconstruction performance of DMSM across different path counts. As observed, increasing the number of paths consistently improves PSNR and SSIM, indicating enhanced reconstruction quality. However, beyond $N=15$, further gains become negligible, leading us to set $N=15$ as the default configuration. A similar trend is observed for uncertainty estimation. Table \ref{tab:uncertainty_quantitative} shows that as the number of paths increases, the PCC between pixel-wise uncertainty estimation and error improves but converges around $N=15$.

\begin{figure}[h]
  \centering
  \includegraphics[width=0.38\textwidth]{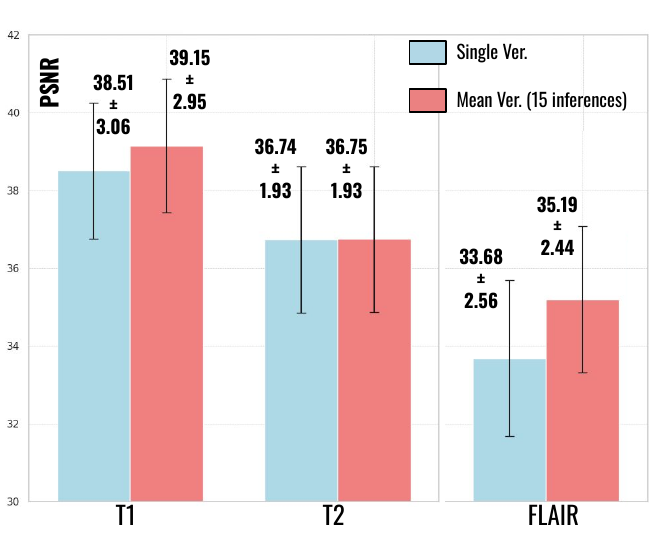}
  \caption{Quantitative analysis on the reconstruction improvement from the multi-path inference. Consistent improvements from multi-path inference (red bar) to single-path inference (blue bar) are found for all sequences on fastMRI dataset under the 4$\times$ setting here. }
  \label{fig:multi-improvement}
\end{figure}

\begin{table} [htb!]
\footnotesize
\centering
\caption{{Ablation study results of DMSM with different numbers of inference paths. Analysis with fastMRI dataset under 4$\times$ acceleration setting is reported. Inference time indicates sampling time for one single reconstructed MRI slice, with the unit of seconds. GFLOPs refers to Giga floating-point operations the model takes for the inference process.}}
\label{tab:multi-number-psnr}
\resizebox{0.5\textwidth}{!}{%
    \begin{tabular}{l|c|c|c|c|c}
        \hline
        \textbf{PSNR/dB}          & T1            & T2              & FLAIR    &Inference Time     &GFLOPs     \Tstrut\Bstrut\\
        \hline
        \hline
        N = 5 & $38.65\pm3.06$   & $36.74\pm1.92$     & $34.35\pm2.58$   &$1.15s$   &$10432.59$  \Tstrut\Bstrut\\
        \hline
        N = 10         & $39.10\pm3.00$   & $36.74\pm1.95$    & $35.10\pm2.49$   &$2.35s$   &$20865.17$ \Tstrut\Bstrut\\
        \hline
        N = 15         & $39.15\pm2.95$   & $36.75\pm1.93$     & $35.19\pm2.44$    &$3.50s$    &$31297.76$\Tstrut\Bstrut\\
        \hline
    \end{tabular}
    }
\end{table}

\begin{figure}[htb!]
\centering
\includegraphics[width=0.45\textwidth]{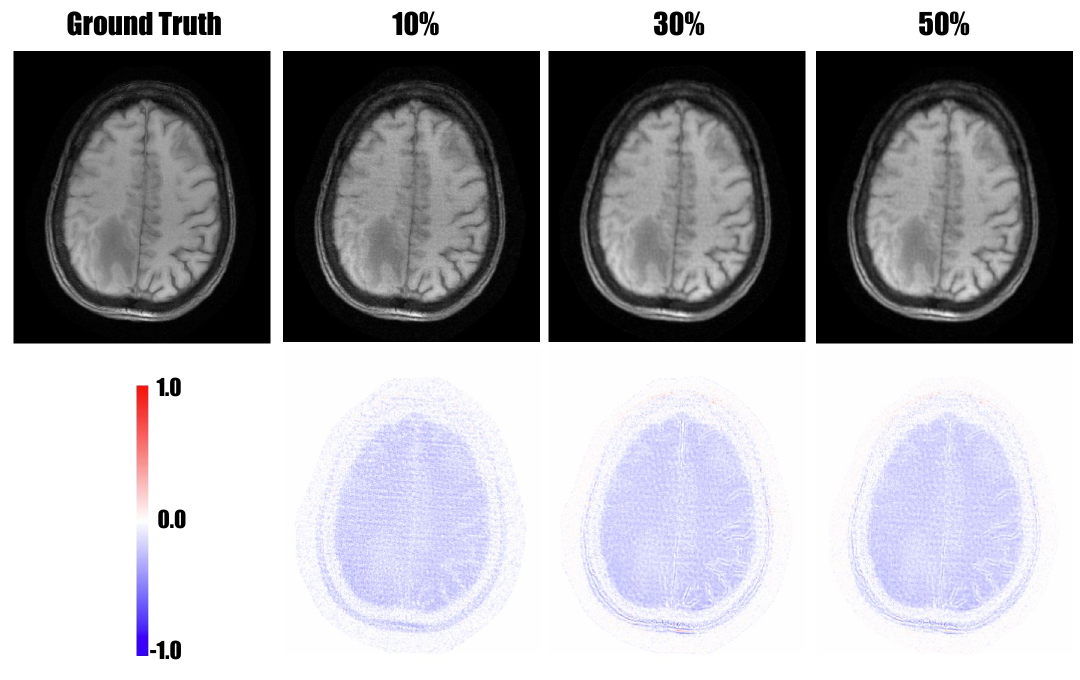}
\caption{{Visualization of ablation studies on partition mask strategy. The DMSM reconstruction (top row) and the corresponding error map (bottom) row are shown. Results on excluding the image-domain loss (2nd column), LHAN (3rd column), and DC layer (4th column) are visualized for a T1 example from fastMRI with 4$\times$ acceleration setting.}}
\label{fig:partition_mask_img}
\end{figure}

\noindent\textbf{{Impact of Partition Mask Strategy:}}
{The partition mask strategy facilitates a self-supervised learning objective by decoupling the training process into three concurrent threads. Beyond its structural role, the mask ratio also impacts reconstruction quality. As summarized in Table \ref{tab:partition_mask_table}, quantitative results across mask ratios (10\%, 30\%, and 50\%) reveal subtle yet consistent performance differences. The 50\% configuration achieves superior reconstruction fidelity, as evidenced by both quantitative metrics and qualitative visualizations in Figure \ref{fig:partition_mask_img}. Error maps further confirm that 50\% yields optimal results, particularly at brain edges and within groove regions.}

\begin{table} [htb!]
\footnotesize
\centering
\caption{{Ablation study results of DMSM with different partition mask ratios. Analysis with fastMRI dataset under 4$\times$ acceleration setting is reported.}}
\label{tab:partition_mask_table}
    \begin{tabular}{l|c|c|c}
        \hline
        \textbf{PSNR/dB}          & T1            & T2              & FLAIR              \Tstrut\Bstrut\\
        \hline
        \hline
        10\%         & $38.16\pm2.97$   & $35.90\pm2.00$     & $34.00\pm2.68$      \Tstrut\Bstrut\\
        \hline
        30\% & $38.89\pm2.95$   & $36.10\pm1.90$     & $34.41\pm2.54$     \Tstrut\Bstrut\\
        \hline
        50\%         & $39.15\pm2.95$   & $36.75\pm1.93$     & $35.19\pm2.44$    \Tstrut\Bstrut\\
        \hline
        
    \end{tabular}
\end{table}

\noindent\textbf{Impact of Loss Weighting Parameter:} 
{We further investigate the impact of the parameter $\lambda_{KC}$ set in training. When changing $\lambda_{KC}$ from 1 to 10, the reconstruction performance shows the best at $\lambda_{KC} = 5$ across all scenarios. The quantitative results are given in Table \ref{tab:loss_weight_table}.}

\begin{table} [htb!]
        \footnotesize
        \centering
        \caption{{Ablation study results of DMSM with different $\lambda_{KC}$ used in training. Analysis with fastMRI dataset under 4$\times$ acceleration setting is reported.}}
        \label{tab:loss_weight_table}
            \begin{tabular}{l|c|c|c}
                \hline
                \textbf{PSNR/dB}          & T1            & T2              & FLAIR              \Tstrut\Bstrut\\
                \hline
                \hline
                $\lambda_{KC} = 1$ & $39.02\pm2.95$   & $35.36\pm1.95$     & $34.79\pm2.52$     \Tstrut\Bstrut\\
                \hline
                $\lambda_{KC} = 5$         & $39.15\pm2.95$   & $36.75\pm1.93$     & $35.19\pm2.44$    \Tstrut\Bstrut\\
                \hline
                $\lambda_{KC} = 10$         & $39.10\pm3.00$   & $36.30\pm2.00$     & $35.10\pm2.45$      \Tstrut\Bstrut\\
                \hline
            \end{tabular}
        \end{table}

\section{Discussion} 


In this work, we introduced DMSM, a lightweight diffusion-based reconstruction model trained using a dual-domain self-supervised strategy. Our approach enables high-quality MRI reconstruction while providing reliable uncertainty estimation. Our dual-domain self-supervised strategy leverages the complementary strengths of the image and k-space domains. By enforcing consistency between these domains, the model inherently respects the physical constraints of MRI acquisition while learning robust priors from undersampled data. Importantly, our approach relies solely on undersampled k-space data for both training and reconstruction, addressing a key limitation in clinical settings where fully sampled data is often unavailable. Building on this self-supervised framework, we developed a lightweight and efficient diffusion model tailored for MRI reconstruction. Our proposed backbone, which integrates a Lightweight Hybrid Attention Network (LHAN) with a Data Consistency (DC) module, enhances reconstruction accuracy while maintaining computational efficiency. Given a randomly noised undersampled MR image as input, LHAN extracts meaningful reconstruction features using attention mechanisms, while DC ensures consistency in the k-space domain with the initial undersampled input. Within LHAN, we introduced a parameter-free symmetric activation function to compute attention maps, replacing traditional heavy self-attention mechanisms used in Transformer-based models. This design significantly reduces trainable parameters compared to baseline diffusion models. Additionally, the cross-attention module facilitates effective guidance from global latent variables (in our case, time-index features) at each diffusion step. Overall, this architecture balances high reconstruction quality with computational efficiency by integrating multi-scale and multi-domain information effectively. Furthermore, we implemented a multi-path inference strategy, which plays a crucial role in improving both reconstruction quality and interpretability. By leveraging a trained network to generate multiple reconstructions with subtle variations, we achieve consistent improvements in quantitative metrics. 
Additionally, uncertainty maps computed from these reconstructions exhibit a strong spatial correlation with error maps, highlighting regions of higher uncertainty. This feature enhances clinical interpretability by directing radiologists' attention to areas that may require closer inspection. {These benefits enhance our network's overall performance and improve the model's explainability, enabling the identification of trustworthy predictions within clinically significant regions. It was also proven that the uncertainty estimation from the diffusion model could be used for further fine-tuning the model to further enhance the performance\cite{chen2025uncertainty}. Notably, while alternative uncertainty estimation techniques exist, such as test-time augmentation (TTA)\cite{wang2019aleatoric}, these methods are generally better suited to image-to-image translation frameworks. Given that our approach operates directly on k-space data, such techniques are not directly applicable. Also, there are some other works that also address the uncertainty estimation\cite{zeevi2025rate}. This dropout strategy during the inference stage could serve to achieve uncertainty estimation, but requires architectural changes.}
Overall, our experimental results demonstrate that DMSM is not only feasible but also highly competitive compared to both supervised and self-supervised reconstruction methods. 



Despite its promising performance, our study has several limitations that warrant further exploration. First, while our dual-domain self-supervised strategy is theoretically applicable to general diffusion models, we validated it only within a conditional diffusion model framework. Future work could extend this approach to newer paradigms, such as the Bridge Diffusion Model \cite{li2023bbdm,su2022dual}. Second, our experiments were conducted on brain MRI datasets with limited contrasts (T1, T2, FLAIR, PD). Although the method is designed to be modality- and anatomy-agnostic, its generalizability to other anatomical regions (e.g., knee MRI) and MRI sequences remains unverified. Future work will explore its applicability to multi-organ and cross-modal reconstruction tasks. Another key limitation is that our evaluation focused primarily on image-level metrics using data from healthy subjects. To enhance clinical relevance, future studies should validate DMSM on pathological cases, such as tumor or lesion reconstruction. Incorporating radiologist assessments and testing on datasets with diverse pathologies will strengthen its potential for real-world deployment. Additionally, while our self-supervised framework eliminates dependence on fully sampled data, integrating a fine-tuning strategy could further improve performance. For instance, initializing the model with weights from a supervised variant (Section \ref{method:implementation and dataset}) and fine-tuning it on unseen undersampled datasets may enhance adaptability to new acquisition protocols and hardware. {On the other hand, there are also recent works on substituting the raw input with the general feature representation of the undersampled MRI, where the feature representation is obtained by contrastive pretraining a feature representation network on a variety of undersampled MRI data \cite{ekanayake2025cl}. Those kinds of add-on components could potentially further improve our model's performance. Finally, the multi-path inference strategy, while beneficial for uncertainty estimation, increases inference time proportionally to the number of paths. With the increase of inference paths, the sampling and processing time of single single slice grows linearly. Considering the large clinical data in real-world scenarios, this could lead to potential challenges of implementation. To mitigate this, future work could explore accelerated inference techniques\cite{salimans2022progressive,lu2022dpm,chen2024accelerating} to reduce computational overhead without sacrificing reconstruction quality.}

\section{Conclusion}
We proposed a Dual-domain Multi-path Self-supervised Diffusion Model (DMSM) for accelerated MRI reconstruction, addressing key limitations of existing methods in fully-sampled data dependency, computational efficiency, and uncertainty estimation. Our method integrates dual-domain self-supervised training, a lightweight hybrid-attention network, and multi-path inference to achieve high-fidelity reconstructions. The experiment results demonstrate that DMSM achieves superior reconstruction quality compared to state-of-the-art supervised and self-supervised methods, with uncertainty maps potentially offering clinically interpretable guidance.

\bibliographystyle{IEEEtran}
\bibliography{bibliography}
\begin{IEEEbiography}[{\includegraphics[width=1in,height=1.25in,clip,keepaspectratio]{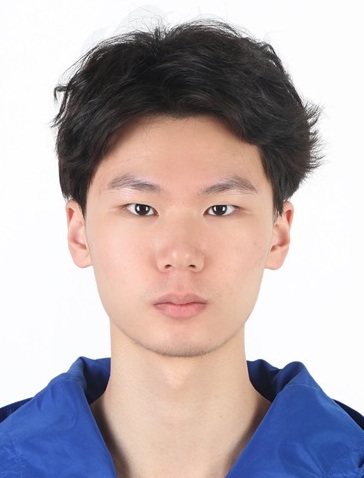}}]{Yuxuan Zhang} received his B.S. in biomedical engineering from Huazhong University of Science and Technology in 2025. He is currently pursuing his M.S. in Computer Science at the University of Southern California. His research interests include AI for medical imaging.
\end{IEEEbiography}
\begin{IEEEbiography}[{\includegraphics[width=1in,height=1.25in,clip,keepaspectratio]{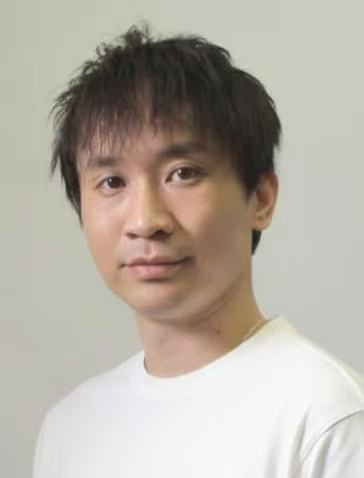}}]{Jinkui Hao} received his Ph.D. from the University of Chinese Academy of Sciences in 2024. He is currently a research fellow at Northwestern University. His research interests focus on medical image processing and AI for healthcare.
\end{IEEEbiography}
\begin{IEEEbiography}[{\includegraphics[width=1in,height=1.25in,clip,keepaspectratio]{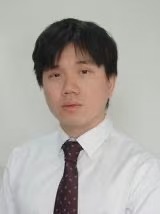}}]{Bo Zhou} received his Ph.D. in Biomedical Engineering from Yale University in 2024. He also received his M.S. in Computer Vision from Carnegie Mellon University in 2018, and another M.S. in Biomedical Engineering from Case Western Reserve University in 2016. He is an Assistant Professor (Research) of Radiology at Northwestern University. He is the IEEE Bruce Hasegawa Young Investigator Medical Imaging Science Awardee with research focuses on general medical AI.
\end{IEEEbiography}

\end{document}